\documentclass[prb,aps,twocolumn,amsmath,superscriptaddress,nofootnoteinbib,nobibnotes]{revtex4-2}

\usepackage{graphicx}
\usepackage{epsfig}
\usepackage{gensymb}
\usepackage{physics}
\begin{document}

\title{Spectrally-modified frequency-swept pulses for optically-driven quantum light sources}

\author{G. R. Wilbur}
\thanks{These two authors contributed equally }
\affiliation{Department of Physics and Atmospheric Science,
Dalhousie University, Halifax, Nova Scotia B3H 4R2 Canada}

\author{A. Binai-Motlagh}
\thanks{These two authors contributed equally }

\affiliation{Department of Physics and Atmospheric Science,
Dalhousie University, Halifax, Nova Scotia B3H 4R2 Canada}

\author{A. Clarke}

\affiliation{Department of Physics and Atmospheric Science,
Dalhousie University, Halifax, Nova Scotia B3H 4R2 Canada}

\author{A. Ramachandran}

\affiliation{Department of Physics and Atmospheric Science,
Dalhousie University, Halifax, Nova Scotia B3H 4R2 Canada}

\author{N. Milson}

\affiliation{Department of Physics and Atmospheric Science,
Dalhousie University, Halifax, Nova Scotia B3H 4R2 Canada}

\author{J. P. Healey}

\affiliation{Department of Physics and Atmospheric Science,
Dalhousie University, Halifax, Nova Scotia B3H 4R2 Canada}

\author{S. O'Neal}

\affiliation{The College of Optics and Photonics, University of Central Florida, Orlando, Florida 32816-2700, USA}
\altaffiliation{Present Address of S. O'Neal: Imec, Kissimmee, Florida, 34744, USA. Present address of D. G. Deppe: SdPhotonics, Richardson, Texas 75081, USA.}

\author{D. G. Deppe}

\affiliation{The College of Optics and Photonics, University of Central Florida, Orlando, Florida 32816-2700, USA}

\author{K. C. Hall}

\affiliation{Department of Physics and Atmospheric Science,
Dalhousie University, Halifax, Nova Scotia B3H 4R2 Canada}
\email[E-mail address: ]{Kimberley.Hall@dal.ca}




\begin{abstract}
We present a driving scheme for solid-state quantum emitters using frequency-swept pulses containing a spectral hole resonant with the optical transition in the emitter.  Our scheme enables high-fidelity state inversion, exhibits robustness to variations in the laser pulse parameters and is immune to phonon-mediated excitation-induced dephasing, benefits that derive from the the insensitivity of the adiabaticity condition to variations in the experimental parameters.  Our resonant driving approach could be combined with spectral filtering of the scattered pump light and photonic devices for enhanced collection efficiency to realize simultaneous high indistinguishability and brightness in single photon source applications.  
\end{abstract}

\maketitle

Deterministic sources of single photons are required for a number of emerging areas in quantum information science and technology, including photonic quantum computing and simulation \cite{Kok:2007,Aspuru_Guzik:2012}, quantum repeaters and networks \cite{Wehner:2018,Borregaard:2020,Kolodynski:2020}, and quantum-enhanced sensing and metrology \cite{Chu:2017,Wang:QDSqueezing2020,Bennett:2016,Muller:2017}.  For such applications, the ideal quantum emitter would be capable of producing a single photon on-demand with successive photons being indistinguishable in all its degrees of freedom.  For solid-state quantum emitters based on semiconductor quantum dots (QDs), considerable improvements in emission properties have been achieved in recent years \cite{Wei:2014,Kuhlmann:TL2015,Somaschi:2016,Ding:2016,Wang:Neartransform2016,Gerhardt:2018,He:2013,Unsleber:2016,Wang:TowardsOptimal2019,Tomm:2021,Huber:2020,Uppu:Onchip2020,Muller:2007,Ates:2009,Liu:2018,Uppu:Scalable2020,Dusanowski:2019,Kirsanske:2017,Thomas:2021,Madsen:2014,Reindl:2019,Heinze:2015,Muller:TPAbiexction2014,Wang:ondemand2019,Schweickert:2018,Chen:highlyEfficient2018,Hanschke:2018,Liu:A_solid_state2019,Huber:TPAGaAs2017,Jayakumar:2013,Jons:2017,Gazzano:2013}, yet the simultaneous achievement of high photon brightness and indistinguishability remains a challenge.  For the highest degree of indistinguishability, resonant pumping of the quantum emitter is ideal since non-resonant excitation can lead to timing jitter due to incoherent relaxation pathways and spectral wandering associated with unwanted excess charges in the vicinity of the emitter \cite{Kuhlmann:TL2015,Somaschi:2016,Gerhardt:2018,He:2013,Kirsanske:2017,Jons:2017,Gazzano:2013,Huber:2015}.  Resonant pumping however necessitates separation of the scattered excitation light from the emitted photon stream.  The resonance fluorescence may be separated from the excitation laser using crossed-polarized excitation and detection \cite{Kuhlmann:TL2015,Somaschi:2016,Ding:2016,Wang:Neartransform2016,Gerhardt:2018,He:2013,Unsleber:2016}, but this approach limits the maximum brightness to 50\% even if all other sources of loss are minimized \cite{Ding:2016,Chen:highlyEfficient2018,Liu:A_solid_state2019,Jons:2017,Huang:Bright2021,Ahn:broadband2021,Claudon:2010,Jurkat:2021,Reimer:2012,Arcari:2014}.  Near-resonant excitation assisted by phonons \cite{Thomas:2021,Madsen:2014,Reindl:2019} enables spectral isolation of the emission from the pump laser but relies on incoherent relaxation processes that reduce coherence.  In two-photon excitation schemes, the pump light and florescence are also nondegenerate due to the nonzero biexciton binding energy \cite{Heinze:2015,Muller:TPAbiexction2014,Wang:ondemand2019,Schweickert:2018,Chen:highlyEfficient2018,Hanschke:2018,Liu:A_solid_state2019,Huber:TPAGaAs2017,Jayakumar:2013}, however the time correlation between the biexciton and exciton emission and the lack of a single polarization tied to the biexciton cascade limit the indistinguishability with this approach \cite{Huber:2013,Yan:Double2021,Wei:Tailoring2021}.   Polarization-sensitive optical microcavities \cite{Wang:TowardsOptimal2019,Tomm:2021} and waveguide modes \cite{Huber:2020,Uppu:Onchip2020,Muller:2007,Ates:2009,Liu:2018,Uppu:Scalable2020,Dusanowski:2019,Kirsanske:2017} have also been utilized in recent years to address the need to separate the excitation light and emitted photoluminescence. The efficiency of excitation and detection is limited in such schemes by the strength of coupling into the associated optical modes.  

A few creative approaches to solving the tradeoff between brightness and indistinguishability have emerged in recent years \cite{He:Coherently2019,Koong:2021,Bracht:2021}.  These approaches rely on the insight that, in the nonlinear driving regime, coherent driving of a two-level system is possible in the absence of light resonant with the optical transition provided that the cancellation of the time-integrated Rabi frequency that normally occurs with off-resonant driving is incomplete. One method uses a bichromatic laser pulse with spectral components on either side of the optical transition but with a node on-resonance \cite{He:Coherently2019,Koong:2021}.  In such a case, the time-integrated pulse area is nonzero provided the symmetry is broken between the red and blue driving conditions, which is naturally caused by phonon-mediated excitation processes and can be enhanced by making the two spectral components have unequal strengths \cite{Koong:2021}.  This approach enables background-free single photon extraction using a simple notch filter resonant with the optical transition, but the maximum inversion is limited to 60\%.  Another pumping scheme referred to as swing-up \cite{Bracht:2021} was recently proposed in which the positive and negative cycles of the Rabi frequency are rendered inequivalent by using two spectral components tuned below the optical transition that beat together.  Excitation of the emitter is possible in this case because the cancellation effects present under non-resonant pumping are incomplete such that the system reaches inversion after many Rabi cycles. For both of these pumping schemes, the final inversion of the emitter is sensitive to the pulse parameters making practical implementation difficult.

Here we show that inversion of a quantum emitter may be achieved using a frequency-swept pulse subjected to spectral filtering, enabling the simultaneous requirements of coherent driving of the emitter and the spectral isolation of the emitted photons from the excitation pulse.  By exploiting adiabatic transfer of the system through an anticrossing in the dressed states of the optically-driven emitter, our scheme is intrinsically robust to fluctuations in the laser pulses and to variations in the optical properties of the quantum emitters often present in solid-state systems.  Furthermore, the use of positively-chirped pulses suppresses phonon-mediated excitation-induced dephasing  \cite{Gustin:2017,Mathew:2014,LukerReview,Wigger:2018,Ramsay_damp:2010,Ramsay_renorm:2010,Vagov:2007} that has been shown to limit the final inversion for many of the above-mentioned excitation schemes \cite{Koong:2021,Bracht:2021}.  We demonstrate our pumping scheme experimentally in a single semiconductor QD, but our approach would be applicable to a wide range of solid-state quantum emitter systems \cite{Atature:2018}.  

The optical excitation scheme presented here is a modified version of adiabatic rapid passage (ARP) which utilizes a frequency-swept pulse to realize quantum state inversion \cite{Loy:1974,Malinovsky:2001,Shore:book}. In ARP, the laser pulse is linearly-chirped, with an electric field given by $E(t)=\frac{1}{2}E_p(t)\exp{[-i(\omega_l t+ \alpha t^2)]}$, where $E_p(t)$ is the pulse envelope, $\omega_l$ is the center frequency of the laser pulse, and $\alpha$ is the temporal chirp.  The instantaneous detuning of the laser frequency from the transition frequency of the quantum emitter ($\omega_0$) is time-dependent, given by $\Delta(t)= \Delta_0-2\alpha t$ with a static detuning of $\Delta_0 = \omega_0$-$\omega_l$, which may be nonzero in the general case, such that the frequency of the pulse is swept from below resonance to above resonance or vice versa depending on the sign of the chirp $\alpha$.  The dressed states of the system in the presence of the light field ($\ket{\Psi_\pm}$) correspond to superpositions of the bare ground and excited states of the emitter ($\ket{0}$ and $\ket{1}$, respectively) with coefficients that evolve in time during the pulse from purely $\ket{0}$ to $\ket{1}$ or vice versa.  The energy splitting between the dressed states is $\sqrt{\Omega(t)^2 + \Delta(t)^2}$, where $\Omega(t)=\frac{\mu E_p(t)}{\hbar}$ is the Rabi frequency and $\mu$ is the dipole moment of the optical transition.    In the limit that $|\Delta \frac{d\Omega}{dt} - \Omega\frac{d\Delta}{dt}| \ll [\Omega^2 +\Delta^2]^{\frac{3}{2}}$ \cite{Shore:book} is satisfied, the control process is adiabatic and the system remains in one of the dressed states while the state itself reverses character, leading to inversion.  The initial state of the system, together with the sign of the pulse chirp, dictates whether the system traverses the anticrossing in $\ket{\Psi_+}$ or $\ket{\Psi_-}$.  

ARP is an intrinsically robust control process because the condition for adiabatic state transfer may be satisfied under a wide range of conditions. The dipole moment and transition frequency of the emitter can vary in ensembles of emitters in solid-state systems due to differences in the environment around each emitter \cite{Atature:2018}.  In addition, the laser system used to drive the emitter can fluctuate in both intensity (leading to changes in the pulse area $\theta = \int_{-\infty}^{\infty}\frac{\mu E_p(t)}{\hbar} dt$) and the laser central frequency $\omega_l$ causing changes in $\Delta_0$.  For ARP, the adiabatic condition may be satisfied in the presence of such variations such that high-fidelity inversion is maintained \cite{Loy:1974,Malinovsky:2001,Shore:book,Ajandetuning}.  Furthermore, the choice of positive chirp in the laser pulses used to drive the emitter leads to inversion via the lower-energy dressed state, which may be used to suppress phonon-induced dephasing processes at low temperature, in contrast to other driving schemes \cite{Mathew:2014}.  Robust inversion has been demonstrated using ARP in solid-state emitters based on excitons in semiconductor QDs in recent years \cite{Ajandetuning,Mathew:2014,Simon:2011,Wu:2011,Gamouras:2013,Kaldewey:2017,Kaldewey_supp:2017,Ramachandran:2020}, including the driving of a triggered single photon source \cite{Wei:2014}.  

In the scheme we present here, the frequency-swept pulse used to drive the quantum emitter are modified by implementing a mask function in the frequency domain that leads to a node on resonance with the optical transition in the emitter.  Such a mask function could be realized by applying a standard dual-mask liquid-crystal array spatial light modulator (SLM) at the focal plane of a 4f pulse shaper configuration \cite{Mathew:2011,Gamouras:2012}.  In this case, the total mask imposed by the shaping system is given by:
\begin{equation}
    M(\omega) = A(\omega)e^{i\Phi(\omega)}
		\label{eqn:mask}
\end{equation}
where $A(\omega) = 1 - e^{-\ln(2)(\omega - \omega_0)^2/\delta^2}$ is the amplitude mask with 2$\delta$ being the full-width-at-half-maximum (FWHM) of the spectral hole. The phase mask is given by $\Phi(\omega) = \frac{\phi^{''}}{2} (\omega -\omega_0)^2$, where the spectral chirp $\phi^{''}$ is related to the temporal chirp $\alpha$ by $\alpha=2\phi''/[\tau_0^4/(2\ln{(2)})^2+(2\phi'')^2]$.  For application of such a spectrally-modified chirped laser pulse to triggered single photon sources, a high quality band-pass filter can be used to block the pump light and pass the emitter fluorescence with near unity efficiency. As we show here, the condition for adiabatic transfer through the anticrossing and concomitant inversion of the quantum emitter may still be achieved in the presence of a spectral hole centered on the emitter transition.  Furthermore this coherent driving scheme, consistent with traditional ARP, enables the suppression of phonon-induced dephasing and is robust to the experimental variations present in real solid-state emitter and laser systems.    

\begin{figure*}[!h]\vspace{0pt}

   \centering
	    \includegraphics[width=13cm]{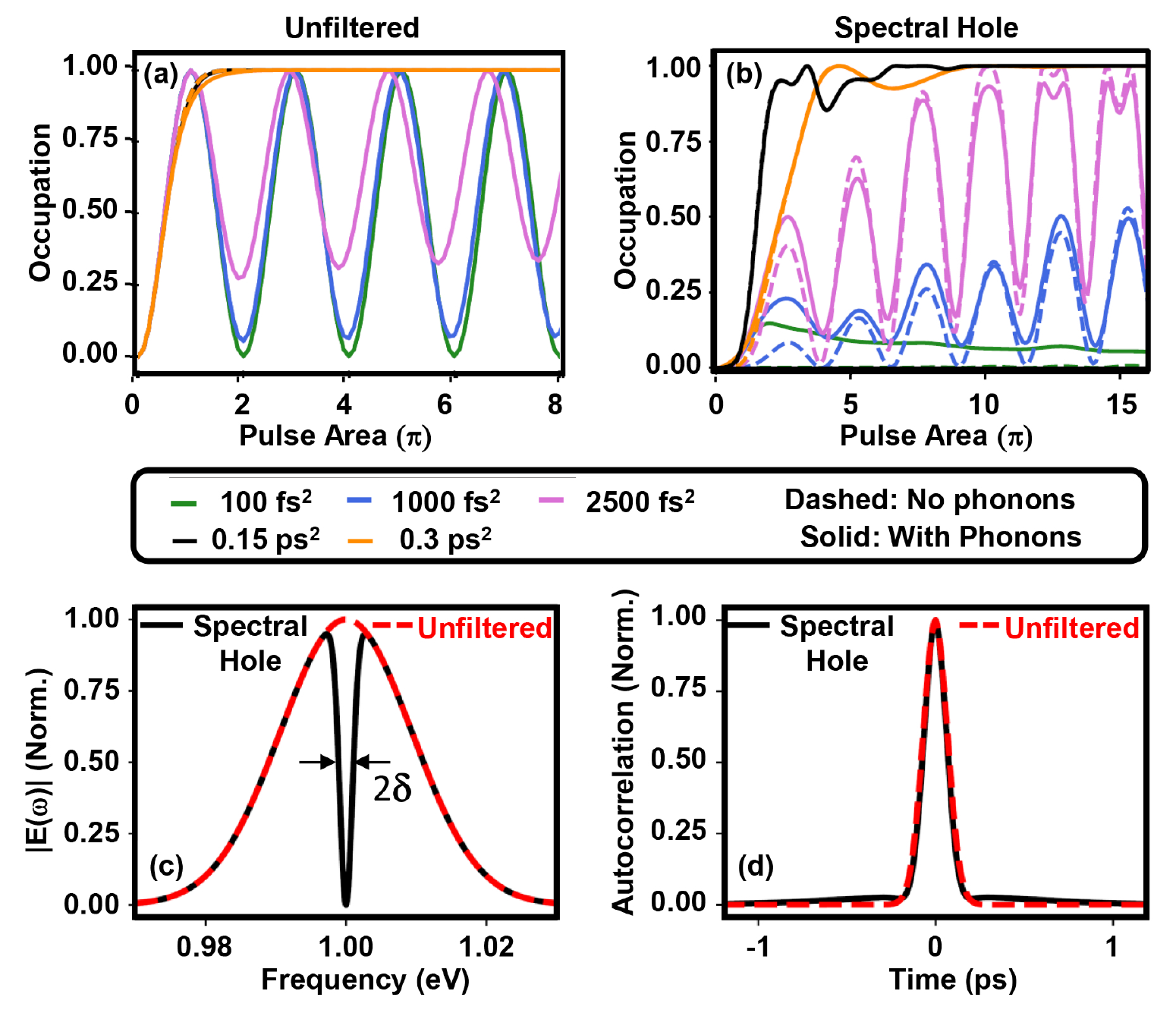}
    \caption{(a) Results of calculations of the quantum state dynamics for optical driving of a single semiconductor QD with an unfiltered Gaussian laser pulse for different values of laser pulse chirp.  (b) Same as (a) for a spectral hole pulse with 2$\delta$~=~2.1~meV.  In (a),(b) the solid (dashed) curves indicate the calculated state dynamics with (without) coupling to LA phonons. (c) Laser spectrum for the Gaussian unfiltered pulse (red dashed curve) and spectral hole pulse (solid black curve).  (d) Calculated pulse autocorrelation corresponding to the spectra in (c).     
}
    \label{fig:Figure1}
\end{figure*}

The results of numerical simulations of the quantum state dynamics illustrating the efficacy of our excitation scheme for quantum state inversion are shown in Fig.~\ref{fig:Figure1}(a) and Fig.~\ref{fig:Figure1}(b).  These simulations were targeted to the situation of an optically-driven single semiconductor QD and used a density matrix approach taking into account deformation potential interactions with acoustic phonons \cite{Ramsay:JAP2011}.  Fig.~\ref{fig:Figure1}(a) and Fig.~\ref{fig:Figure1}(b) show the occupation of the excited state as a function of pulse area for a range of values of positive pulse chirp for an unfiltered Gaussian spectrum [(a)] and a pulse subject to spectral filtering using the mask in Eqn.~\ref{eqn:mask} with 2$\delta$~=~2.1~meV [(b)], referred to as a spectral hole pulse.  The corresponding pulse characteristics are shown in Fig.~\ref{fig:Figure1}(c) and Fig.~\ref{fig:Figure1}(d).  For the unfiltered pulse, a transition from the Rabi rotation regime to the adiabatic regime is observed as $\phi^{''}$ is increased, consistent with previous work \cite{Wei:2014,Ajandetuning,Mathew:2014,Simon:2011,Wu:2011,Gamouras:2013,Kaldewey:2017,Kaldewey_supp:2017,Ramachandran:2020}.  Once the adiabatic regime is reached, which occurs for $\phi^{''}\gtrsim$0.1~ps$^{2}$ and $\theta \gtrsim$2$\pi$, the upper state occupation is insensitive to changes in pulse area, a key signature of adiabatic rapid passage and robust inversion.  For the spectral hole pulse with $\phi^{''}$~=~0, the occupation vanishes in the absence of phonon coupling because the time-integrated Rabi frequency is zero, as discussed previously for the case of unchirped pulses \cite{Koong:2021}.  As shown in Fig.~\ref{fig:Figure1}(b), the inclusion of phonons leads to a small but nonzero state occupation because phonons breaks the spectral symmetry of the driving conditions through the introduction of incoherent dynamics \cite{Koong:2021}.  The results in Fig.~\ref{fig:Figure1}(b) show that the use of frequency-swept pulses also serves to break the symmetry:  The state occupation increases with increasing $\phi^{''}$.  For low $\phi^{''}$, the occupation versus pulse area resembles an imperfect Rabi rotation.  For $\phi^{''}$ exceeding 0.1~ps$^{2}$, the adiabatic regime is reached.  In this limit, for sufficiently large pulse area, the inversion process is robust, for both the unfiltered pulse and the spectral hole pulse.

\begin{figure*}[!h]\vspace{0pt}
\centering
    \includegraphics[width=14cm]{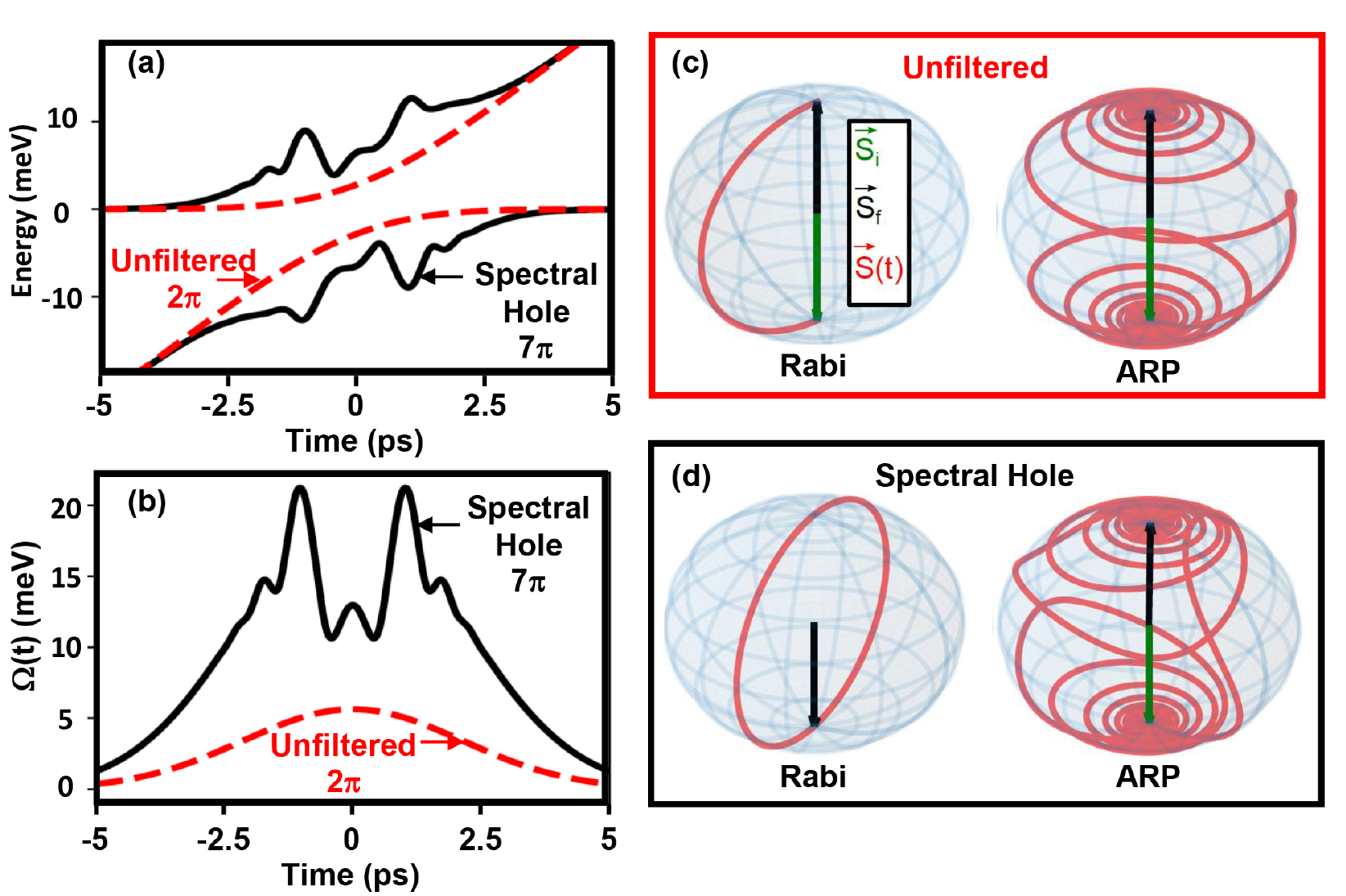}
    \caption{(a)  Energies $E_{\pm}$ of the dressed states $\ket{\Psi_{\pm}}$ for an unfiltered Gaussian pulse (dashed red curve) and the spectral hole pulse with 2$\delta$~=~2.1~meV (solid black curve) for $\phi^{''}$~=~0.15~ps$^{2}$.    (b) Calculated Rabi frequency as a function of time corresponding to the dressed state energies in (a). The pulse areas in (a) and (b) were chosen as the minimum value leading to robust inversion, corresponding to 2$\pi$ (7$\pi$) for the unfiltered (spectral hole) pulse. (c)  Quantum state dynamics depicted on the Bloch sphere for an unfiltered Gaussian pulse with $\phi^{''}$~=~0 and a pulse area of $\theta$~=~$\pi$ (left) and $\phi^{''}$~=~0.15~ps$^{2}$ with $\theta$~=~3.4$\pi$ (right).  (d) Bloch spheres for a spectral hole pulse with 2$\delta$~=~2.1~meV for $\phi^{''}$~=~0 and a pulse area of $\theta$~=~3.4$\pi$ (left) and $\phi^{''}$~=~0.15~ps$^{2}$ with $\theta$~=~3.4$\pi$ (right).
}
    \label{fig:Figure2}
\end{figure*}

We can gain insight into the quantum state dynamics from the energies of the dressed states of the optically-driven system ($E_{\pm}$), which are shown in Fig.~\ref{fig:Figure2}(a).  Robust inversion occurs for frequency-swept pulses with or without a spectral hole because in both cases, the adiabatic condition for quantum state transfer is satisfied.  The introduction of the spectral hole leads to structure in the temporal shape of the pulse, as shown in Fig.~\ref{fig:Figure2}(b), which has the effect of increasing the threshold pulse area for ARP.  Calculations as a function of the hole width ($\delta$) indicate that the primary cause for this increase is the transfer of energy into the wings of the pulse, which reduces the size of the Rabi frequency at time t~=~0 thereby reducing the magnitude of the dressed-state splitting at the anticrossing.  The adiabatic condition $|\Delta \frac{d\Omega}{dt} - \Omega\frac{d\Delta}{dt}| \ll [\Omega^2 +\Delta^2]^{\frac{3}{2}}$ is then recovered for larger $\theta$, with a threshold pulse area that increases with increasing $\delta$ (See Supplemental Information). Nevertheless, for a given value of $\delta$ and sufficiently large pulse area, robust inversion occurs.  The corresponding quantum state dynamics using a Bloch vector representation are shown without phonon coupling in Fig.~\ref{fig:Figure2}(c) and Fig.~\ref{fig:Figure2}(d).  For excitation with the spectral hole pulse with $\phi^{''}$~=~0, the initial and final states of the quantum emitter coincide with the ground state and no net excitation occurs due to the zero time-integrated Rabi frequency.  The inclusion of phonons causes only a slight deviation of the final Bloch vector from the south pole.   For chirped pulse excitation, the characteristic adiabatic rapid passage Bloch vector trajectory reflecting the time-dependent detuning is observed and robust inversion of the quantum emitter is realized for either the unfiltered or spectral hole pulse, with only a slight modification of the trajectory in the latter case.    

The results in Fig. 1(b) indicate that the fractional change in the occupation resulting from the impact of phonons diminishes as the chirp is increased.  This reflects the use of positively chirped pulses, which suppresses LA phonon-mediated transitions between the dressed states within the adiabatic regime at low temperatures \cite{Mathew:2011}.  In Fig.~\ref{fig:Figure1}(b), the dashed and solid curves coincide for $\phi^{''}$ and $\theta$ above the threshold for ARP, indicating that phonons have no influence on the quantum state dynamics in this limit.  Our results therefore indicate that all of the advantages of ARP for optical inversion of quantum emitter systems, including the ability to suppress decoherence tied to phonons and the robustness of the inversion process to fluctuations in the laser source, are maintained in the presence of a spectral hole resonant with the optical transition in the emitter.

\begin{figure*}[!h]\vspace{0pt}
\centering
    \includegraphics[width=15cm]{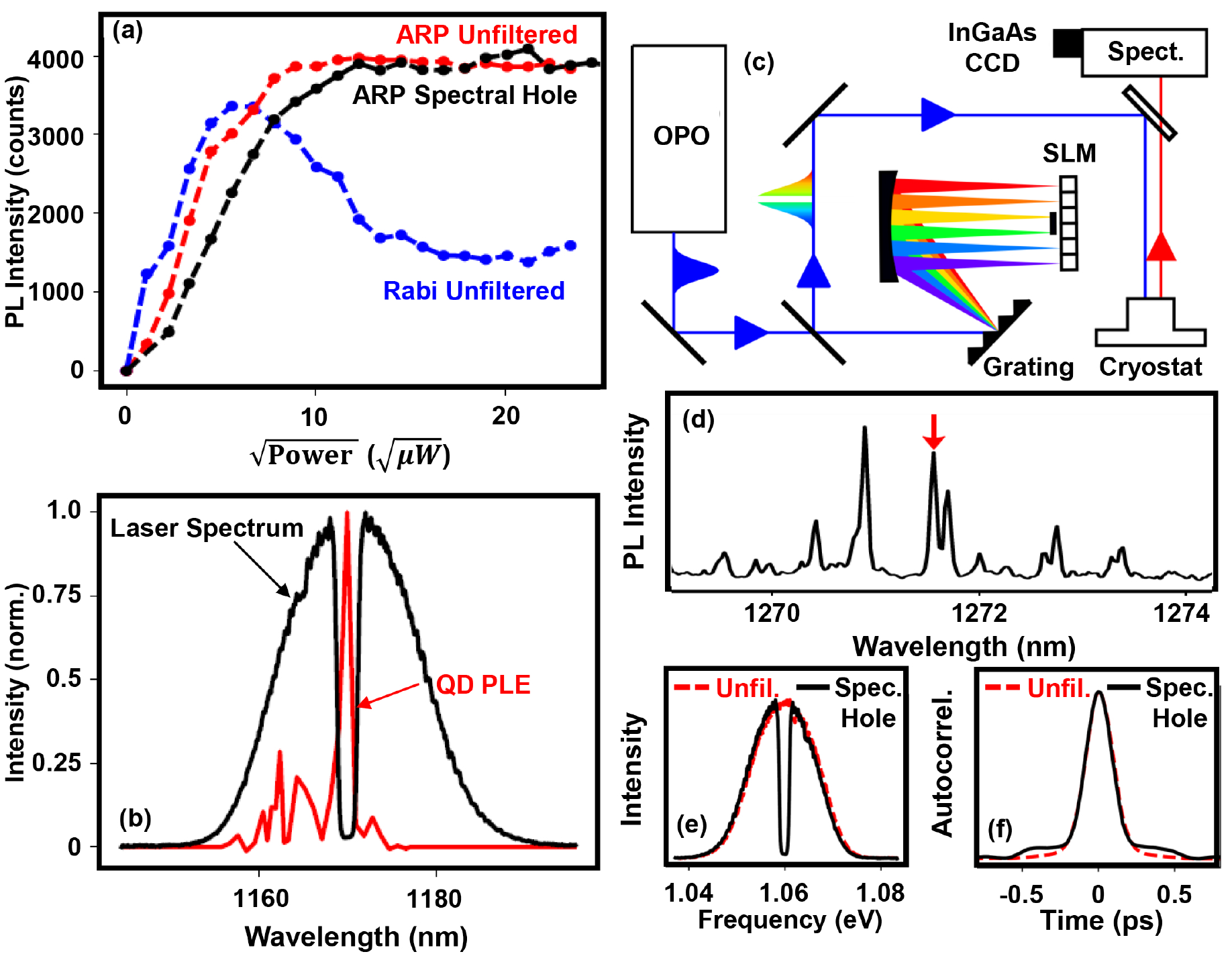}
    \caption{(a) Measured PL intensity from a single semiconductor QD under optical driving by an unfiltered pulse with $\phi^{''}$~=~0 (blue symbols) and $\phi^{''}$~=~0.15~ps$^{2}$ (red symbols), and for a spectral hole pulse with a 3~meV hole and $\phi^{''}$~=~0.15~ps$^{2}$ (black symbols).  (b) Laser spectrum for the spectral hole pulse (black curve) together with the photoluminescence excitation (PLE) spectrum of the QD used in the experiments (red curve). (c) Schematic diagram of the quantum control apparatus. (d) Micro-PL showing the s-shell emission from individual QDs. The arrow indicates the QD used for the experiments in (a).  (e) Measured pulse autocorrelation for the unfiltered pulse (red dashed curve) and the spectral hole pulse (black solid curve).  (f) Laser spectrum for the unfiltered pulse (red dashed curve) and the spectral hole pulse (black solid curve).   
}
    \label{fig:Figure3}
\end{figure*}

We demonstrate our optical driving scheme experimentally in a single InGaAs QD.  The results of these experiments are shown in Fig.~\ref{fig:Figure3}.  The laser source was a tunable infrared optical parametric oscillator with a pulse duration after dispersion compensation of 110~fs.  The experimental demonstration of our excitation scheme was carried out on the p-shell optical transition in the QD, with detection of the final quantum state via the s-shell photoluminescence emission (PL), although our scheme could also be used for ground state pumping as was assumed in the theoretical calculations in Fig.~\ref{fig:Figure1} and Fig.~\ref{fig:Figure2}.  The laser was tuned to the p-shell transition in the QD centered at 1170 nm (Fig.~\ref{fig:Figure3}(b)).  The s-shell PL centered at 1271.6~nm (arrow in Fig.~\ref{fig:Figure3}(d)) was detected using a 0.75~m spectrometer equipped with a liquid nitrogen-cooled InGaAs array detector.  The QD sample was housed in a cryostat equipped with an attocube nanopositioner.  A 4f pulse shaper was used to introduce dispersion compensation and to impose the desired spectral chirp.  Experiments were carried out using the dual mask function of the SLM enabling simultaneous amplitude and phase shaping as well as using a pure phase mask together with a physical blocker in front of the SLM, with similar results.  The data in Fig.~\ref{fig:Figure3}(a) were taken using a pure phase mask and a physical blocker that produced a hole of width 3~meV. The laser pulse spectrum and autocorrelation with and without the spectral hole are shown in Fig.~\ref{fig:Figure3}(e) and Fig.~\ref{fig:Figure3}(f). 

The PL intensity versus the square root of the excitation power, which is proportional to the pulse area, is shown in Fig.~\ref{fig:Figure3}(a) for an unfiltered pulse with zero chirp (blue symbols), for an unfiltered pulse with $\phi^{''}$~=~0.15~ps$^{2}$ (red symbols) and for the spectral hole pulse with $\phi^{''}$~=~0.15~ps$^{2}$ (black symbols).  For the unfiltered pulse with zero chirp, a damped Rabi rotation is observed.  For excitation by the chirped, unfiltered pulse the PL intensity saturates at a constant value for pulse areas above the threshold for ARP, as observed in prior experiments on similar QDs \cite{Wei:2014,Ajandetuning,Mathew:2014,Simon:2011,Wu:2011,Gamouras:2013,Kaldewey:2017,Kaldewey_supp:2017,Ramachandran:2020}.  For the chirped spectral hole pulse, the exciton inversion also exhibits a saturation behaviour.  For chirped pulse excitation, the pulse area required to reach full inversion is approximately 30\% larger for the spectral hole pulse than the unfiltered pulse. These findings are in qualitative agreement with the theoretical predictions in Fig.~\ref{fig:Figure1}(b).  

For single photon source applications, a variety of experimental approaches have been used to eliminate the tradeoff between indistinguishability and brightness, which together require resonant driving of the two-level system in the quantum emitter and the efficient separation of the emitted photon stream from the scattered pump light.  The optical driving scheme we present here, which relies on spectrally-modified frequency-swept pulses, may be used in conjunction with a simple notch filter resonant with the optical transition in the quantum emitter to simultaneously satisfy these two requirements, setting the stage for high-performance single photon sources.  Our pumping scheme offers the significant advantage over phonon-assisted pumping \cite{Glassl:2013,Ardelt:2014}, bichromatic driving \cite{He:Coherently2019,Koong:2021} and swing up \cite{Bracht:2021} schemes that LA phonon-mediated excitation-induced dephasing does not limit the maximum occupation at low temperature when positively-chirped pulses are used.  In addition unlike the bichromatic or swing-up schemes, our approach is insensitive to the details of the driving pulse (pulse area, magnitude of chirp, width of spectral hole).  Our findings indicate that a spectral hole width as large as several meV may be used to invert the emitter.  This would enable the parallel excitation of multiple emitters with unequal transition energies as long as these transitions fall within the filtered band \cite{Ajandetuning}, adding versatility for application of our approach in real solid-state emitter systems. 

In summary, we present an optical driving scheme for solid-state quantum emitter systems that utilizes spectrally-modified, frequency-swept laser pulses.  Our findings indicate that the adiabatic regime for quantum state inversion may still be reached in the presence of a spectral hole coincident with the optical transition in the quantum emitter with only a modest increase in pulse area relative to traditional ARP.  We find that this scheme retains all of the benefits of ARP, including insensitivity to fluctuations in the driving laser pulse and the ability to suppress phonon-mediated dephasing processes.  We demonstrate our approach experimentally in a single InGaAs semiconductor QD.  The use of the driving scheme we present here in conjunction with a simple notch filter to separate the emitted photon stream from the excitation pulse, together with photonic cavity geometries to enhance collection efficiency, would enable the realization of quantum emitters with simultaneous high photon indistinguishability and brightness.  Our findings will support the development of high-performance, optically-driven solid-state quantum emitters for applications such as photonic quantum computing and quantum networks.

\section*{Funding}
This research is supported by the Natural Sciences and Engineering Research Council
of Canada.

\section*{Disclosures}
The authors declare no conflicts of interest.


\newpage
\newpage

\newpage
\newpage
\newpage

%

\newpage
\newpage
\newpage
\newpage

\begin{thebibliography}{1}





\bibitem{Kok:2007} P. Kok, W. J. Munro, K. Nemoto, T. C. Ralph, J. P. Dowling, and G. J. Milburn, \textit{Linear optical quantum computing with photonic qubits}, Rev. Mod. Phys. {\bf 79}, 135 (2007).

\bibitem{Aspuru_Guzik:2012}  A. Aspuru-Guzik and P. Walther, \textit{Photonic quantum simulators}, Nature Physics {\bf 8}, 285 (2012).



\bibitem{Wehner:2018} S. Wehner, D. Elkouss, R. Hanson, \textit{Quantum internet: A vision for the road ahead}, Science {\bf 362}, 303 (2018). 

\bibitem{Borregaard:2020} J. Borregaard, H. Pichler, T. Schr\"{o}der, M. D. Lukin, P. Lodahl, and A. S. S\o{}rensen, \textit{One-way quantum repeater based on near-deterministic photon-emitter interfaces}, Phys. Rev. X {\bf 10}, 021071 (2020).

\bibitem{Kolodynski:2020} J. Ko\l{}ody\'{n}ski, A. M\'{a}ttar, P. Skrzypczyk, E. Woodhead, D. Cavalcanti, K. Banaszek, and A. Ac\'{i}n, Quantum {\bf 4}, 260 (2020).



\bibitem{Chu:2017} X.-L. Chu, S. G\"{o}tzinger, V. Sandoghdar, \textit{A single molecule as a high-fidelity photon gun for producing intensity-squeezed light}, Nature Photonics {\bf 11}, 58 (2017).

\bibitem{Wang:QDSqueezing2020} H. Wang, J. Qin, S. Chen, M.-C. Chen, X. You, X. Ding, Y.-H. Huo, Y. Yu, C. Schneider, S. H\"{o}fling, M. Scully, C.-Y. Lu, and J.-W. Pan, \textit{Observation of intensity squeezing in resonance fluorescence from a solid state device}, Phys. Rev. Lett. {\bf 125}, 153601 (2020).


\bibitem{Bennett:2016} A. J. Bennett, J. P. Lee, D. J. P. Ellis, T. Meany, E. Murray, F. F. Floether, J. P. Griffiths, I. Farrer, D. A. Richie, A. J. Shields, \textit{Cavity-enhanced coherent light scattering from a quantum dot}, Science Advances {\bf 2}, 1501256 (2016).

\bibitem{Muller:2017} M. M\"{u}ller, H. Vural, C. Schneider, A. Rastelli, O. G. Schmidt, S. H\"{o}fling, and P. Michler, \textit{Quantum-dot single-photon sources for enhanglement-enhanced interferometry}, Phys. Rev. Lett. {\bf 118}, 257402 (2017).




\bibitem{Wei:2014} Y.-J. Wei, Y.-M. He, M.-C. Chen, Y.-N. Hu, Y. He, D. Wu, C. Schneider, M. Kamp, S. H\"{o}fling, C.-Y. Lu, and J.-W. Pan, \textit{Deterministic and robust generation of single photons from a single quantum dot with 99.5\% indistinguishability using adiabatic rapid passage}, Nano Lett. {\bf 14}, 6515 (2014).


\bibitem{Kuhlmann:TL2015} A. V. Kuhlmann, J. H. Prechtel, J. Houel, A. Ludwig, D. Reuter, A. D. Wieck, and R. J. Warburton, \textit{Transform-limited single photons from a single quantum dot}, Nat.  Commun. {\bf 6}, 8204 (2015).

\bibitem{Somaschi:2016} N. Somaschi, V. Giesz, L. De Santis, J. C. Loredo, M. P. Almeida, G. Hornecker, S. L. Portalupi, T. Grange, C. Ant\'{o}n, J. Demory, C. G\'{o}mez, J. Sagnes, N. D. Lanzillotti-Kimura, A. Lema\'{i}tre, A. Auffeves, A. G. White, L. Lanco, and P. Senellart, \textit{Near-optimal single-photon sources in the solid state}, Nature Photonics {\bf 10}, 340 (2016). 

\bibitem{Ding:2016} X. Ding, Y. He, Z.-C. Duan, N. Gregersen, M.-C. Chen, S. Unsleber, S. Maier, C. Schneider, M. Kamp, S. H\"{o}fling, C.-Y. Lu, and J.-W. Pan, \textit{On-demand single photons with high extraction efficiency and near-unity indistinguishability from a resonantly driven quantum dot in a micropillar}, Phys. Rev. Lett. {\bf 116}, 020401 (2016).  

\bibitem{Wang:Neartransform2016} H. Wang, Z.-C. Duan, Y.-H. Li, S. Chen, J.-P. Li, Y.-M. He, M.-C. Chen, Y. He, X. Ding, C.-Z. Peng, C. Schneider, M. Kamp, S. H\"{o}fling, C.-Y. Lu, and J.-W. Pan, \textit{Near-transform-limited single photons from an efficient solid-state quantum emitter}, Phys. Rev. Lett. {\bf 116}, 213601 (2016).

\bibitem{Gerhardt:2018}  S. Gerhardt, J. Iles-Smith, D. P. S. McCutcheon, Y.-M. He, S. Unsleber, S. Betzold, N. Gregersen, J. M\o{}rk, S. H\"{o}fling, and C. Schneider, \textit{Intrinsic and environmental effects on the interference properties of a high-performance quantum dot single-photon source}, Phys. Rev. B {\bf 97}, 195432 (2018). 

\bibitem{He:2013} Y.-M. He, Y. He, Y.-J. Wei, D. Wu, M. Atat\"{u}re, C. Schneider, S. H\"{o}fling, M. Kamp, C.-Y. Lu, and J.-W. Pan, \textit{On-demand semiconductor single-photon source with near-unity indistinguishability}, Nat. Nanotechnol. {\bf 8}, 213 (2013).

\bibitem{Unsleber:2016} S. Unsleber, Y.-M. He, S. Gerhardt, S. Maier, C.-Y. Lu, J.-W. Pan, N. Gregersen, M. Kamp, C. Schneider, and S. H\"{o}fling, \textit{Highly indistinguishable on-demand resonance fluorescence photons from a deterministic quantum dot micropillar device with 74\% extraction efficiency}, Optics Express {\bf 24}, 8539 (2016).


\bibitem{Wang:TowardsOptimal2019} H. Wang, Y.-M. He, T.-H. Chung, H. Hu, Y. Yu, S. Chen, X. Ding, M.-C. Chen, J. Qin, X. Yang, R.-Z. Liu, Z.-C. Duan, J.-P. Li, S. Gerhardt, K. Winkler, J. Jurkat, L.-J. Wang, N. Gregersen, Y.-H. Huo, Q. Dai, S. Yu, S. H\"{o}fling, C.-Y. Lu, and J.-W. Pan, \textit{Towards optimal single-photon sources from polarized microcavities}, Nat. Photonics {\bf 13}, 770 (2019).
 
\bibitem{Tomm:2021} N. Tomm, A. Javadi, N. O. Antoniadis, D. Najer, M. C. L\"{o}bl, A. R. Korsch, R. Schott, S. R. Valentin, A. D. Wieck, A. Ludwig, and R. J. Warburton, \textit{A bright and fast source of coherent single photons}, Nat. Nanotechnol. {\bf 16}, 399 (2021).


\bibitem{Huber:2020} T. Huber, M. Davanco, M. M\"{u}ller, Y. Shuai, O. Gazzano, and G. S. Solomon, \textit{Filter-free single-photon quantum dot resonance fluorescence in an integrated cavity-waveguide device}, Optica {\bf 7}, 380 (2020). 

\bibitem{Uppu:Onchip2020} R. Uppu, H. T. Eriksen, H. Thyrrestrup, A. D. U\u{g}urlu, Y. Wang, S. Scholz, A. D. Wieck, A. Ludwig, M. C. L\"{o}bl, R. J. Warburton, P. Lodahl, and L. Midolo, \textit{On-chip deterministic operation of quantum dots in dual-mode waveguides for a plug-and-play single-photon source}, Nat. Commun. {\bf 11}, 3782 (2020). 

\bibitem{Muller:2007} A. Muller, E. B. Flagg, P. Bianucci, X. Y. Wang, D. G. Deppe, W. Ma, J. Zhang, G. J. Salamo, M. Xiao, and C. K. Shih, \textit{Resonance fluorescence from a coherently driven semiconductor quantum dot in a cavity}, Phys. Rev. Lett. {\bf 99}, 187402 (2007). 

\bibitem{Ates:2009} S. Ates, S. M. Ulrich, S. Reitzenstein, A. L\"{o}ffler, A. Forchel, and P. Michler, \textit{Post-selected indistinguishable photons from the resonance fluorescence of a single quantum dot in a microcavity}, Phys. Rev. Lett. {\bf 103}, 167402 (2009). 

\bibitem{Liu:2018} F. Liu, A. J. Brash, J. O'Hara, L. M. P. P. Martins, C. L. Phillips, R. J. Coles, B. Royall, E. Clarke, C. Bentham, N. Prtljaga, I. E. Itskevich, L. R. Wilson, M. S. Skolnick, and A. M. Fox. \textit{High Purcell factor generation of indistinguishable on-chip single photons}, Nat. Nanotech. {\bf 13}, 835 (2018).


\bibitem{Uppu:Scalable2020} R. Uppu, F. T. Pedersen, Y. Wang, C. T. Olesen, C. Papon, X. Zhou, L. Midolo, S. Scholz, A. D. Wieck, A. Ludwig, P. Lodahl, \textit{Scalable integrated single-photon source}, Sci. Adv. {\bf 6}, eabc8268 (2020).

\bibitem{Dusanowski:2019} L. Dusanowski, S.-H. Kwon, C. Schneider, and S. H\"{o}fling, \textit{Near-unity indistinguishability single photon source for large-scale integrated quantum optics}, Phys. Rev. Lett. {\bf 122}, 173602 (2019).

\bibitem{Kirsanske:2017} G. Kir\u{s}ansk\.{e}, H. Thyrrestrup, R. S. Daveau, C. L. Dreeßen, T. Pregnolato, L. Midolo, P. Tighineanu, A. Javadi, S. Stobbe, R. Schott, A. Ludwig, A. D. Wieck, S. I. Park, J. D. Song, A. V. Kuhlmann, I. S\"{o}llner, M. C. L\"{o}bl, R. J. Warburton, and P. Lodahl, \textit{Indistinguishable and efficient single photons from a quantum dot in a planer nanobeam waveguide}, Phys. Rev. B {\bf 96}, 165306 (2017).


\bibitem{Thomas:2021} S. E. Thomas, M. Billard, N. Coste, S. C. Wein, Priya, H. Ollivier, O. Krebs, L. Taza\"{i}rt, A. Harouri, A. Lemaitre, I. Sagnes, C. Anton, L. Lanco, N. Somaschi, J. C. Loredo, and P. Senellart, \textit{Bright polarized single-photon source based on a linear dipole}, Phys. Rev. Lett. {\bf 126}, 233601 (2021).

\bibitem{Madsen:2014} K. H. Madsen, S. Ates, L. Liu, A. Javadi, S. M. Albrecht, I. Yeo, S. Stobbe, and P. Lodahl, \textit{Efficient out-coupling of high-purity single photons from a coherent quantum dot in a photonic crystal cavity}, Phys. Rev. B {\bf 90}, 155303 (2014).

\bibitem{Reindl:2019} M. Reindl, J. H. Weber, D. Huber, C. Schimpf, S. F. Covre da Silva, S. L. Portalupi, R. Trotta, P. Michler, and A. Rastelli, \textit{Highly indistinguishable single photons from incoherently excited quantum dots}, Phys. Rev. B {\bf 100}, 155420 (2019).


\bibitem{Heinze:2015} D. Heinze, D. Breddermann, A. Zrenner, and S. Schumacher, \textit{A quantum dot single-photon source with on-the-fly all-optical polarization control and timed emission}, Nat. Commun. {\bf 6}, 8473 (2015).

\bibitem{Muller:TPAbiexction2014} M. M\"{u}ller, S. Bounouar, K. D. J\"{o}ns, M. Gl\"{a}ssl, and P. Michler, \textit{On-demand generation of indistinguishable polarization-entangled photon pairs}, Nature Photonics {\bf 8}, 224 (2014).

\bibitem{Wang:ondemand2019} H. Wang, H. Hu, T.-H. Chung, J. Qin, X. Yang, J.-P. Li, R.-Z. Liu, H.-S. Zhong, Y.-M. He, X. Ding, Y.-H. Deng, Q. Dai, Y.-H. Huo, S. H\"{o}fling, C.-Y. Lu, and J.-W. Pan, \textit{On-demand semiconductor source of entangled photons which simultaneously has high fidelity, efficiency, and indistinguishability}, Phys. Rev. Lett. {\bf 122}, 113602 (2019).

\bibitem{Schweickert:2018} L. Schweickert, K. D. J\"{o}ns, K. D. Zeuner, S. F. Covre da Silva, H. Huang, T. Lettner, M. Reindl, J. Zichi, R. Trotta, A. Rastelli, and V. Zwiller, \textit{On-demand generation of background-free single photons from a solid-state source}, Appl. Phys. Lett. {\bf 112}, 093106 (2018).

\bibitem{Chen:highlyEfficient2018} Y. Chen, M. Zopf, R. Keil, F. Ding, and O. G. Schmidt, \textit{Highly-efficient extraction of entangled photons from quantum dots using a broadband optical antenna}, Nat. Commun. {\bf 9}, 2994 (2018).

\bibitem{Hanschke:2018} L. Hanschke, K. A. Fischer, S. Appel, D. Lukin, J. Wierzbowski, S. Sun, R. Trivedi, J. Vu\u{c}kovi\'{c}, J. J. Finley, and K. M\"{u}ller, \textit{Quantum dot single-photon sources with ultra-low multi-photon probability}, npj Quantum Information {\bf 4}, 43 (2018).

\bibitem{Liu:A_solid_state2019} J. Liu, R. Su, Y. Wei, B. Yao, S. F. Covre da Silva, Y. Yu, J. Iles-Smith, K. Srinivasan, A. Rastelli, J. Li, and X. Wang, \textit{A solid-state source of strongly entangled photon pairs with high brightness and indistinguishability}, Nat. Nanotechnol. {\bf 14}, 586 (2019). 

\bibitem{Huber:TPAGaAs2017} D. Huber, M. Reindl, Y. Huo, H. Huang, J. S. Wildmann, O. G. Schmidt, A. Rastelli, and R. Trotta, \textit{Highly indistinguishable and strongly entangled photons from symmetric GaAs quantum dots}, Nat. Commun. {\bf 8}, 15506 (2017).

\bibitem{Jayakumar:2013} H. Jayakumar, A. Predojevi\'{c}, T. Huber, T. Kauten, G. S. Solomon, and G. Weihs, \textit{Deterministic photon pairs and coherent optical control of a single quantum dot}, Phys. Rev. Lett. {\bf 110}, 135505 (2013).

\bibitem{Jons:2017} K. D. J\"{o}ns, L. Schweickert, M. A. M. Versteegh, D. Dalacu, P. J. Poole, A. Gulinatti, A. Giudice, V. Zwiller, and M. E. Reimer, \textit{Bright nanoscale source of deterministic entangled photon pairs violating Bell's inequality}, Sci. Rep. {\bf 7}, 1700 (2017).


\bibitem{Gazzano:2013} O. Gazzano, S. Michaelis de Vasconcellos, C. Arnold, A. Nowak, E. Galopin, I. Sagnes, L. Lanco, A. Lemaitre and P. Senellart, \textit{Bright solid-state sources of indistinguishable single photons}, Nat. Commun. {\bf 4}, 1425 (2013).


\bibitem{Huber:2015} T. Huber, A. Predojevi\'{c}, D. F\"{o}ger, G. Solomon, and G. Weihs, \textit{Optimal excitation conditions for indistinguishable photons from quantum dots}, New. J. Phys. {\bf 17}, 123025 (2015).



\bibitem{Huang:Bright2021} H. Huang, S. Manna, C. Schimpf, M. Reindl, X. Yuan, Y. Zhang, S. F. Covre da Silva, A. Rastelli, \textit{Bright single photon emission from quantum dots embedded in a broadband planar optical antenna}, Advanced Optical Materials {\bf 9}, 2001490 (2021). 

\bibitem{Ahn:broadband2021} D. H. Ahn, Y. D. Jang, J. S. Baek, C. Schneider, S. H\"{o}fling, D. Lee, \textit{A broad-band planar-microcavity quantum-dot single-photon source with a solid immersion lens}, Appl. Phys. Lett. {\bf 118}, 174001 (2021). 

\bibitem{Claudon:2010} J. Claudon, J. Bleuse, N. S. Malik, M. Bazin, P. Jaffrennou, N. Gregersen, C. Sauvan, P. Lalanne, J.-M. G\'{e}rard, \textit{A highly efficient single-photon source based on a quantum dot in a photonic nanowire}, Nature Photonics {\bf 4}, 174 (2010).

\bibitem{Jurkat:2021} J. Jurkat, M. Mocza\l{}a-Dusanowska, M. A. Jacobsen, A. Predojevi\'{c}, T. Huber, N. Gregersen, S. H\"{o}fling, and C. Schneider, \textit{Technological implementation of a photonic Bier-Glas cavity}, Phys. Rev. Materials {\bf 5}, 064603 (2021).

\bibitem{Reimer:2012} M. E. Reimer, G. Bulgarini, N. Akopian, M. Hocevar, M. B. Bavinck, M. A. Verheijen, E. P. A. M. Bakkers, L. P. Kouwenhoven, and V. Zwiller, \textit{Bright single-photon sources in bottom-up tailored nanowires}, Nat. Commun. {\bf 3}, 737 (2012).

\bibitem{Arcari:2014} M. Arcari, I. S\"{o}llner, A. Javadi, S. Lindskov Hansen, S. Mahmoodian, J. Liu, H. Thyrrestrup, E. H. Lee, J. D. Song, S. Stobbe, and P. Lodahl, \textit{Near-unity coupling efficiency of a quantum emitter to a photonic crystal waveguide}, Phys. Rev. Lett. {\bf 113}, 093603 (2014).


\bibitem{Huber:2013} T. Huber, A. Predojevi\'{c}, H. Zoubi, H. Jayakumar, G. S. Solomon, G. Weihs, \textit{Measurement and modification of biexciton-exciton time correlations}, Optics Express {\bf 21} 9890 (2013).

\bibitem{Yan:Double2021} J. Yan, S. Liu, X. Lin, Y. Ye, J. Yu, L. Wang, Y. Yu, Y. Zhao, Y. Meng, X. Hu, D. Wang, C. Jin, F. Liu, \textit{Double-pulse generation of indistinguishable single photons with optically controlled polarization}, arXiv:2109.09279 (2021).


\bibitem{Wei:Tailoring2021} Y. Wei, S. Liu, X. Li, Y. Yu, X. Su, S. Li, S. Xiang, H. Liu, H. Hao, H. Ni, S. Yu, Z. Niu, J. Liu, X. Wang, \textit{Tailoring solid-state single-photon sources with stimulated emissions}, arXiv:2109.09284 (2021).


\bibitem{He:Coherently2019} Y.-M. He, H. Wang, C. Wang, M.-C. Chen, X. Ding, J. Qin, Z.-C. Duan, S. Chen, J.-P. Li, R.-Z. Liu, C. Schneider, M. Atat\"{u}re, S. H\"{o}fling, C.-Y. Lu, and J.-W. Pan, \textit{Coherently driving a single quantum two-level system with dichromatic laser pulses}, Nature Physics {\bf 15}, 941 (2019).

\bibitem{Koong:2021} Z. X. Koong, E. Scerri, M. Rambach, M. Cygorek, M. Brotons-Gisbert, R. Picard, Y. Ma, S. I. Park, J. D. Song, E. M. Gauger, B. D. Gerardot, \textit{Coherent dynamics in quantum emitters under dichromatic excitation}, Phys. Rev. Lett. {\bf 126}, 047403 (2021).

\bibitem{Bracht:2021} T. K. Bracht, M. Cosacchi, T. Seidelmann, M. Cygorek, A. Vagov, V. M. Axt, T. Heindel, D. E. Reiter, \textit{Swing-up of quantum emitter population using detuned pulses}, PRX Quantum 2, 040354 (2021).



\bibitem{Gustin:2017} C. Gustin and S. Hughes, \textit{Influence of electron-phonon scattering for an on-demand quantum dot single-photon source using cavity-assisted adiabatic passage} Phys. Rev. B {\bf 96}, 085305 (2017).

\bibitem{Mathew:2014} R. Mathew, E. Dilcher, A. Gamouras, A. Ramachandran, H. Y. S. Yang, S. Freisem, D. Deppe and K. C. Hall, \textit{Subpicosecond adiabatic rapid passage on a single semiconductor quantum dot: Phonon-mediated dephasing in the strong-driving regime}, Phys. Rev. B {\bf 90}, 035316 (2014).


\bibitem{LukerReview} S. Luker and D. E. Reiter, \textit{A review on optical excitation of semiconductor quantum dots under the influence of phonons}, Semicond. Sci. Technol. {\bf 34}, 063002 (2019).

\bibitem{Wigger:2018} D. Wigger, C. Schneider, S. Gerhardt, M. Kamp, S. Hofling, T. Kuhn, and J. Kasprzak, \textit{Rabi oscillations of a quantum dot exciton coupled to acoustic phonons: coherence and population readout}, Optica {\bf 5}, 1442 (2018).

\bibitem{Ramsay_damp:2010} A. J. Ramsay, A. V. Gopal, E. M. Gauger, A. Nazir, B. W. Lovett, A. M. Fox, and M. S. Skolnick, \textit{Damping of exciton Rabi rotations by acoustic phonons in optically excited InGaAs/GaAs quantum dots}, Phys. Rev. Lett. {\bf 104}, 017402 (2010).

\bibitem{Ramsay_renorm:2010} A. J. Ramsay, T. M. Godden, S. J. Boyle, E. M. Gauger, A. Nazir, B. W. Lovett, A. M. Fox, and M. S. Skolnick, \textit{Phonon-induced Rabi-frequency renormalization of optically driven single InGaAs/GaAs quantum dots}, Phys. Rev. Lett. {\bf 105}, 177402 (2010).


\bibitem{Vagov:2007} A. Vagov, M. D. Croitoru, V. M. Axt, T. Kuhn, and F. M. Peeters, \textit{Nonmonotonic field dependence of damping and reappearance of Rabi oscillations in quantum dots}, Phys. Rev. Lett. {\bf 98}, 227403 (2007).

\bibitem{Atature:2018} M. Atature, D. Englund, N. Vamivakas, S.-Y. Lee, and J. Wrachtrup, \textit{Material platforms for spin-based photonic quantum technologies}, Nature Reviews Materials {\bf 3}, 38 (2018). 


\bibitem{Loy:1974} M. M. T. Loy, \textit{Observation of population inversion by optical adiabatic rapid passage}, Phys. Rev. Lett. {\bf 32}, 814 (1974).

\bibitem{Malinovsky:2001} V. S. Malinovsky and J. L. Krause, \textit{General theory of population transfer by adiabatic rapid passage with intense, chirped laser pulses}, Eur. Phys. J. D {\bf 14}, 147 (2001).

\bibitem{Shore:book} Bruce W. Shore, \textit{Manipulating quantum structures using laser pulses},  (Cambridge University Press, New York, 2011), pp 513-521.

\bibitem{Ajandetuning} A. Ramachandran, J. Fraser-Leach, S. O.'Neal, D. G. Deppe, and K. C. Hall, \textit{Experimental quantification of the robustness of adiabatic rapid passage for quantum state inversion in semiconductor quantum dots}, Optics Express {\bf 29}, 41766 (2021).

\bibitem{Mathew:2014} R. Mathew, E. Dilcher, A. Gamouras, A. Ramachandran, H. Y. S. Yang, S. Freisem, D. Deppe and K. C. Hall, \textit{Subpicosecond adiabatic rapid passage on a single semiconductor quantum dot: Phonon-mediated dephasing in the strong-driving regime}, Phys. Rev. B {\bf 90}, 035316 (2014).


\bibitem{Simon:2011} C.-M. Simon, T. Belhadj, B. Chatel, T. Amand, P. Renucci, A. Lemaitre, O. Krebs, P. A. Dalgarno, R. J. Warburton, X. Marie, and B. Urbaszek, \textit{Robust quantum dot exciton generation via adiabatic rapid passage with frequency-swept optical pulses}, Phys. Rev. Lett. {\bf 106}, 166801 (2011).

\bibitem{Wu:2011} Y. Wu, I. M. Piper, M. Ediger, P. Brereton, E. R. Schmidgall, P. R. Eastham, M. Hugues, M. Hopkinson, and R. T. Phillips, \textit{Population inversion in a single InGaAs quantum dot using the method of adiabatic rapid passage}, Phys. Rev. Lett. {\bf 106}, 067401 (2011).

\bibitem{Gamouras:2013} A. Gamouras, R. Mathew, S. Freisem, D. G. Deppe, and K. C. Hall, \textit{Simultaneous deterministic control of distant qubits in two semiconductor quantum dots}, Nano Lett. {\bf 13}, 4666 (2013).

\bibitem{Kaldewey:2017} T. Kaldewey, S. Luker, A. V. Kuhlmann, S. R. Valentin, A. Ludwig, A. D. Wieck, D. E. Reiter, T. Kuhn, and R. J. Warburton, \textit{Coherent and robust high-fidelity generation of a biexciton in a quantum dot by rapid adiabatic passage}, Phys. Rev. B {\bf 95}, 161302(R) (2017).

\bibitem{Kaldewey_supp:2017} T. Kaldewey, S. Luker, A. V. Kuhlmann, S. R. Valentin, J.-M. Chauveau, A. Ludwig, A. D. Wieck, D. E. Reiter, T. Kuhn, and R. J. Warburton, \textit{Demonstrating the decoupling regime of the electron-phonon interaction in a quantum dot using chirped pulse optical excitation}, Phys. Rev. B {\bf 95}, 241306(R) (2017).

\bibitem{Ramachandran:2020} A. Ramachandran, G. R. Wilbur, S. O.'Neal, D. G. Deppe, and K. C. Hall, \textit{Suppression of decoherence tied to electron-phonon coupling in telecom-compatible quantum dots: low-threshold reappearance regime for quantum state inversion}, Optics Letters {\bf 45}, 6498 (2020).

\bibitem{Gamouras:2012} A. Gamouras, R. Mathew, and K. C. Hall, \textit{Optically engineered ultrafast pulses for controlled rotations of exciton qubits in semiconductor quantum dots}, J. Appl. Phys. {\bf 112}, 014313 (2012).

\bibitem{Mathew:2011} R. Mathew, C. E. Pryor, M. E. Flatte, and K. C. Hall, \textit{Optimal quantum control for conditional rotation of exciton qubits in semiconductor quantum dots}, Phys. Rev. B {\bf 84}, 205322 (2011). 

\bibitem{Ramsay:JAP2011} A. J. Ramsay, T. M. Godden, S. J. Boyle, E. M. Gauger, A. Nazir, B. W. Lovett, A. V. Gopal, A. M. Fox, and M. S. Skolnick, \textit{Effect of detuning on the phonon induced dephasing of optically driven InGaAs/GaAs quantum dots}, J. Appl. Phys. {\bf 109} 102415 (2011).

\bibitem{Glassl:2013} M. Gl\"{a}ssl, A. M. Barth, and V. M. Axt, \textit{Proposed robust and high-fidelity preparation of excitons and biexcitons in semiconductor quantum dots making active use of phonons}, Phys. Rev. Lett. {bf 110}, 147401 (2013).

\bibitem{Ardelt:2014} P.-L. Ardelt, L. Hanschke, K. A. Fischer, K. M\"{u}ller, A. Kleinkauf, M. Koller, A. Bechtold, T. Simmet, J. Wierzbowski, H. Riedl, G. Abstreiter, and J. J. Finley, \textit{Dissipative preparation of the exciton and biexciton in self assembled quantum dots on picosecond time scales}, Phys. Rev. B {\bf 90}, 241404(R) (2014).

\end{thebibliography}
\end{document}